\documentclass[journal]{IEEEtran}

\newcommand{\tf}[1]{\textcolor{red}{\uline{#1}}}

\newcommand{\E}{\ensuremath{\mathbb{E}}}
\newcommand{\I}{\ensuremath{\mathbb{I}}}
\newcommand{\Entr}{\ensuremath{\mathbb{H}}}

\newcommand{\Rbmd}{\ensuremath{\text{R}_{\text{BMD}}}\xspace}

\newcommand{\Rsym}{\ensuremath{\text{R}_{\text{SYM}}}\xspace}

\newcommand{\X}{\ensuremath{X}\xspace}
\newcommand{\Xn}{\ensuremath{X^n}\xspace}
\newcommand{\Y}{\ensuremath{Y}\xspace}
\newcommand{\Yn}{\ensuremath{Y^n}\xspace}
\newcommand{\B}{\ensuremath{\boldsymbol{B}}\xspace}
\newcommand{\setX}{\ensuremath{\mathcal{X}}\xspace}
\newcommand{\x}{\ensuremath{x}\xspace}
\newcommand{\y}{\ensuremath{y}\xspace}

\newcommand{\momforth}{\standmom{4}\xspace}
\newcommand{\momsixth}{\standmom{6}\xspace}
\newcommand{\kth}{\ensuremath{k^{\text{th}}}\xspace}
\newcommand{\auxch}{\ensuremath{q_{\Y|\X}}\xspace}

\newcommand{\Ptx}{\ensuremath{P_{\text{tx}}}\xspace}

\newcommand{\NoiseEff}{\ensuremath{\sigma^2_{\text{eff}}}\xspace}

\newcommand{\NoiseNLI}{\ensuremath{\sigma^2_{\text{NLI}}}\xspace}

\newcommand{\chiforth}{\ensuremath{\chi_{4}}\xspace}
\newcommand{\chisixth}{\ensuremath{\chi_{6}}\xspace}
\newcommand{\chinomom}{\ensuremath{\chi_{0}}\xspace}
\newcommand{\standmom}[1]{\ensuremath{\hat{\mu}_{#1}}\xspace}

\newcommand{\tento}[1]{$\times$10\textsuperscript{#1}}

\usepackage[utf8]{inputenc}

\usepackage{amsmath,amssymb}

\usepackage{xcolor}
\usepackage{graphicx}

\usepackage{cite}
\usepackage{xspace}
\usepackage[normalem]{ulem}
\usepackage[pdftex,hidelinks,bookmarks=false,citecolor=blue,urlcolor=blue]{hyperref} 

\allowdisplaybreaks[4] 

\newcommand{\ASE}{\ensuremath{\sigma^2_{\text{ASE}}}\xspace}
\newcommand{\SNR}{\ensuremath{\text{SNR}_{\text{eff}}}\xspace}

\renewcommand{\tf}[1]{#1}

\begin{document}
\title{On Probabilistic Shaping of Quadrature Amplitude Modulation for the Nonlinear Fiber Channel}

\author{Tobias~Fehenberger,~\IEEEmembership{Student Member,~IEEE}, Alex Alvarado,~\IEEEmembership{Senior Member,~IEEE},\\Georg Böcherer,~\IEEEmembership{Member,~IEEE}, and~Norbert~Hanik,~\IEEEmembership{Senior Member,~IEEE}
\thanks{Tobias Fehenberger, Georg Böcherer, and Norbert Hanik are with the Institute for Communications Engineering, Technical University of Munich (TUM), 80333 Munich, Germany (\mbox{Emails:}~tobias.fehenberger@tum.de, georg.boecherer@tum.de, norbert.hanik@tum.de).} 
\thanks{Alex Alvarado is with the Optical Networks Group, University College London (UCL), London, WC1E 7JE, UK (\mbox{Email:}~alex.alvarado@ieee.org).}
\thanks{Alex Alvarado's research is supported by the Engineering and Physical Sciences Research Council (EPSRC) project UNLOC (EP/J017582/1), United Kingdom. Georg Böcherer's research is supported by the German Federal Ministry of Education and Research in the framework of an Alexander von Humboldt Professorship.}
}

\maketitle
\begin{abstract}
 Different aspects of probabilistic shaping for a multi-span optical communication system are studied. First, a numerical analysis of the additive white Gaussian noise (AWGN) channel investigates the effect of using a small number of input probability mass functions (PMFs) for a range of signal-to-noise ratios (SNRs), instead of optimizing the constellation shaping for each SNR. It is shown that if a small penalty of at most 0.1~dB SNR to the full shaping gain is acceptable, just two shaped PMFs are required per quadrature amplitude modulation (QAM) over a large SNR range. For a multi-span wavelength division multiplexing (WDM) optical fiber system with 64QAM input, it is shown that just one PMF is required to achieve large gains over uniform input for distances from 1,400~km to 3,000~km. \tf{Using recently developed theoretical models that extend the Gaussian noise (GN) model} and full-field split-step simulations, we illustrate the ramifications of probabilistic shaping on the effective SNR after fiber propagation. Our results show that, \tf{for a fixed average optical launch power}, a shaping gain is obtained for the noise contributions from fiber amplifiers and modulation-independent nonlinear interference (NLI), whereas shaping simultaneously causes a penalty as it leads to an increased NLI. However, this nonlinear shaping loss is found to have a relatively minor impact, and optimizing the shaped PMF with a modulation-dependent GN model confirms that the PMF found for AWGN is also a good choice for a multi-span fiber system.
\end{abstract}


\IEEEpeerreviewmaketitle

\section{Introduction}
	\IEEEPARstart{T}{hrough} a series of revolutionary technological advances, optical transmission systems have enabled the growth of Internet traffic for decades \cite{Richardson2010Science_CapacityCrunch}. Most of the huge bandwidth of fiber systems is in use \cite{Essiambre2012ProcIEEE_CapacityTrendsReview} and the capacity of the optical core network cannot keep up with the traffic growth \cite{Bayvel2016PhilTransRSocA_MaximizingCapacityReview}.

	The usable bandwidth of an optical communication system with legacy standard single-mode fiber (SMF) is effectively limited by the loss profile of the fiber and the erbium-doped fiber amplifiers (EDFAs) placed between every span. It is thus of high practical importance to increase the spectral efficiency (SE) in optical fiber systems. Even with new fibers, the transceiver will eventually become a limiting factor in the pursuit of higher SE because the practically achievable signal-to-noise ratio (SNR) can be limited by transceiver electronics \cite{Maher2015ECOC_GMICurves}. Digital signal processing (DSP) techniques that are robust against fiber nonlinearities and also offer sensitivity and SE improvements in the linear transmission regime are thus of great interest.

	\begin{figure*}[t]
		\vspace*{-15pt}
		\centering
		\includegraphics{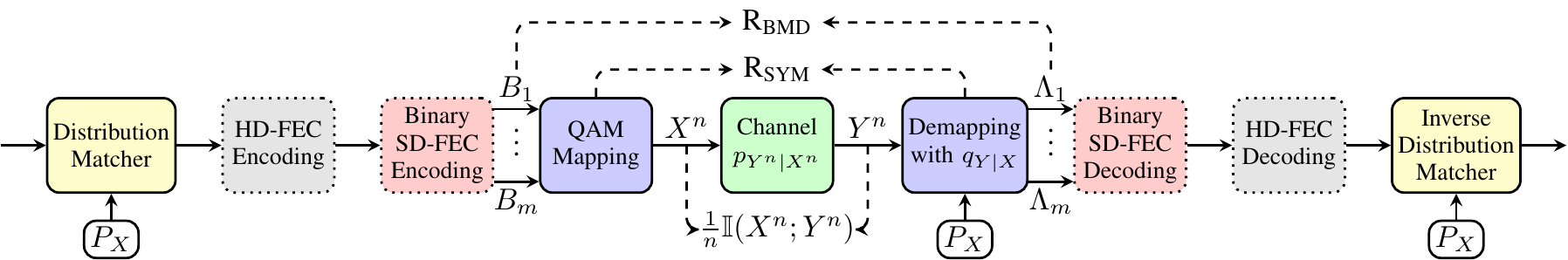}
		\caption{Block diagram of a coded modulation transmitter with probabilistic shaping, a bit-wise demapper that uses the auxiliary channel \auxch, and a binary decoder. The dotted FEC blocks are omitted in the fiber simulations as we focus on AIRs. The transceiver blocks that use the nonuniform distribution are marked with $P_{\X}$.}
		\label{fig:model}
		\vspace{-10pt}
	\end{figure*}

	A technique that fulfills these requirements and that has been very popular in recent years is \textit{signal shaping}.
	There are two types of shaping: geometric and probabilistic.
	In geometric shaping, a nonuniformly spaced constellation with equiprobable symbols is used, whereas in probabilistic shaping, the constellation is on a uniform grid with differing probabilities per constellation point. Both techniques offer an SNR gain up to the ultimate shaping gain of 1.53~dB for the additive white Gaussian noise (AWGN) channel \cite[Sec.~IV-B]{Forney1984JSEL_Shaping}, \cite[Sec.~VIII-A]{Wachsmann1999TransIT_Shaping}. Geometric shaping has been used in fiber optics to demonstrate increased SE \cite{Djordjevic2010OFC_GeometricShapingIPM,Batshon2010PTL_GeometricShapingIPQ,Lotz2013JLT_GeoShaping,EstaranZibar2013ECOC_GeometricShaping,LiuDjordjevic2014_GeometricShapingOSCD,Shiner2014OptExp_8DX,Geller2016JLT_ShapingOverTime}. Probabilistic shaping has attracted considerable attention in fiber optics \cite{Smith2012JLT_CodedModulation,Beygi2014JLT_ShellMappingShaping,Yankov2014PTL_ProbabilisticShaping,Fehenberger2015OFC_ProbShaping,Buchali2015ECOC_ProbShapingExp,DinizDarliMello2016OFC_ProbShapingNetwork,Fehenberger2016PTL_MismatchedShaping,Buchali2016JLT_ProbShapingExp,Yankov2016JLT_ShapingExperiment}. In particular, \cite{Fehenberger2015OFC_ProbShaping,Buchali2015ECOC_ProbShapingExp,Fehenberger2016PTL_MismatchedShaping,Buchali2016JLT_ProbShapingExp} use the probabilistic amplitude-shaping scheme of \cite{Boecherer2015TransComm_ProbShaping} that allows forward-error correction (FEC) to be separated almost entirely from shaping by concatenating a distribution matcher \cite{Schulte2016TransIT_DistributionMatcher} and an off-the-shelf systematic FEC encoder.

	Probabilistic shaping offers several advantages over geometric shaping. Using the scheme in \cite{Boecherer2015TransComm_ProbShaping}, the labeling of the quadrature amplitude modulation (QAM) symbols can remain an off-the-shelf binary reflected Gray code, which gives large achievable information rates (AIRs) for bit-wise decoders and makes exhaustive numerical searching for an optimal labeling obsolete. A further feature of probabilistic shaping that, for fiber-optics, has only been considered in \cite{Buchali2015ECOC_ProbShapingExp,Buchali2016JLT_ProbShapingExp} is that it can yield rate adaptivity, i.e., the overall coding overhead can be changed without modifying the actual FEC. Probabilistic shaping also gives larger shaping gains than purely geometric shaping \cite[Fig.~4.8 (bottom)]{Alvarado2015BICM_Book} for a constellation with a fixed number of points. Given these advantages, we restrict our analysis in this work to probabilistic shaping on a symbol-by-symbol basis. Shaping over several time slots has been studied theoretically \cite{Dar2014ISIT_nonlinearShaping} and is beyond the scope of the present study.

	In this paper, we extend our previous work on probabilistic shaping for optical back-to-back systems \cite{Fehenberger2016PTL_MismatchedShaping} and investigate the impact of shaping for QAM formats on the nonlinear interference (NLI) of an optical fiber channel with wavelength division multiplexing (WDM). For the analysis, we use a \tf{recently developed} modulation-dependent Gaussian noise (GN) model \cite{Dar2013OptExp_PropertiesNLIN} in addition to full-field split-step Fourier method (SSFM) simulations. This GN model includes the impact of the channel input on the NLI by taking into account higher-order standardized moments of the modulation, which allows us to study the impact of probabilistic shaping on the NLI from a theoretical point of view.

	The contributions of this paper are twofold. Firstly, we show that one shaped QAM input, optimized for the AWGN channel, gives large shaping gains also for a multi-span fiber system. This allows potentially for a simplified implementation of probabilistic shaping because just one input PMF can be used for different fiber parameters. Secondly, no significant additional shaping gain is obtained for such a multi-span system with 64QAM when the PMF is optimized to the optical fiber channel using a GN model. The relevance of this result is that numerical optimizations of the channel input \tf{PMF} are shown to be obsolete for many practical long-haul fiber systems.



\section{Fundamentals of Probabilistic Shaping}\label{sec:shaping}

	In the following, we review the basic principles of probabilistic shaping. The focus is on AIRs rather than bit-error ratios after FEC. Both symbol-wise AIRs and AIRs for bit-wise decoding are discussed. For a more detailed comparison, we refer the reader to \cite[Sec.~III]{Buchali2016JLT_ProbShapingExp}, \cite[Ch.~4]{Alvarado2015BICM_Book},\cite{Alvarado2015JLT_BitWise},\cite{Boecherer2016Arxiv_BMDRates}.

\subsection{Achievable Information Rates}

	Consider an independent and identically distributed (iid) discrete channel input $\Xn=\X_1,\X_2,\dots,\X_n$ and the corresponding continuous outputs $\Yn=\Y_1,\Y_2,\dots,\Y_n$. The channel is described by the channel transition probability density $p_{\Yn|\Xn}$, as shown in the center of Fig.~\ref{fig:model}.
	The symbol-wise inputs $\X$ are complex QAM symbols that take on values in $\setX=\{\x_1,\ldots,\x_{M}\}$ according to the probability mass function (PMF) $P_{\X}$ on $\setX$. Without loss of generality, the channel input is normalized to unit energy, i.e., $\E[|X|^2]=1$. The constellation size $|\setX|$ is the modulation order and denoted by $M$. Unless otherwise stated, we consider QAM input that can be decomposed into its constituent one-dimensional (1D) pulse amplitude modulation (PAM) constellation without loss of information. This means that every QAM symbol can be considered as two consecutive PAM symbols that represent the real and imaginary parts of the QAM symbol. The probability of each two-dimensional (2D) QAM constellation is the product of the respective 1D PAM probabilities, denoted by $P_{\text{1D}}$. The analysis in this work is conducted with 2D QAM symbols; it is explicitly stated when a 1D input is considered, which is done mainly for the ease of notation and graphical representation.

	The mutual information (MI) between the channel input and output sequences, normalized by the sequence length, is defined as
	\begin{equation}\label{eq:defMI_sequence}
	  \frac{1}{n} \I(\Xn;\Yn) = \frac{1}{n} \E \left[ \log_2 \frac{p_{\Yn|\Xn}(\Yn|\Xn)}{p_{\Yn}(\Yn)} \right],
	\end{equation}
	where $\E[\cdot]$ denotes expectation and $p_{\Yn}$ is the marginal distribution of $\Yn$. The MI in \eqref{eq:defMI_sequence} is an AIR for a decoder that uses soft metrics based on $p_{\Yn|\Xn}$.


	Since the optical channel is not known in closed form, we cannot directly evaluate \eqref{eq:defMI_sequence}. A technique called mismatched decoding \cite{Ganti2000TransIT_MismatchedDecoding,Secondini2013JLT_AIR} is used in this paper, which gives an AIR for a decoder that operates with the auxiliary channel $q_{\Yn|\Xn}$ instead of the true $p_{\Yn|\Xn}$. In this paper we consider memoryless auxiliary channels of the form
	\begin{equation} \label{eq:auxchannel_memoryless}
		q_{\Yn|\Xn}(\y^n|\x^n) = \prod_{i=1}^{n}q_{\Y_i|\X_i}(\y_i|\x_i),
	\end{equation}
	which means that, in the context of fiber-optics, all correlations over polarization and time are neglected at the decoder. We assume a fixed auxiliary channel, i.e., $q_{\Y_i|\X_i}=q_{\Y|\X} ~ \forall i$, and restrict the analysis in this paper to 2D \tf{circularly symmetric} Gaussian distributions
	\begin{equation} \label{eq:auxchannel}
		q_{\Y|\X}(\y|\x) = \frac{1}{\sqrt{2\pi\sigma^2}} e^{ -\frac{|\y-\x|^2}{2 \sigma^2} },
	\end{equation}
	where $\sigma^2$ is the noise variance of the auxiliary channel, $\x \in \setX$, and \y is complex. For details on the impact of higher-dimensional Gaussian auxiliary channels, see \cite{Fehenberger2016OFC_AIR4DMI,	Eriksson2016JLT_4DMI}. Irrespective of the particular choice of the auxiliary channel, we get a lower bound to $\I(\X;\Y)$ by using $q_{\Y|\X}$ instead of $p_{\Yn|\Xn}$ \cite[Sec.~VI]{Arnold2006TransIT_AchievableRates},
	\begin{align}\label{eq:mismatcheddecoderMI}
		\frac{1}{n}\I(\X^n;\Y^n)  
		& \geq \E \left[ \log_2 \frac{q_{\Y|\X}(\Y|\X)}{q_{\Y}(\Y)} \right] \triangleq \Rsym,
	\end{align}
	where the expectation is taken with respect to $p_{XY}$, and $q_{\Y}(\Y)=\sum_{\x^{\prime} \in \setX} q_{\Y|\X}(\Y|\x^{\prime}) P_X(\x^{\prime})$. The value of \Rsym can be estimated from Monte Carlo simulations of $N$ input-output pairs $(\x_k,\y_k)$ of the channel as
	\begin{equation}\label{eq:_symbMIMonteCarlo}
	  \Rsym \approx \frac{1}{N} \sum_{k=1}^{N} \log_2 \frac{q_{\Y|\X} (\y_k|\x_k)}{q_{\Y}(\y_k)}.
	\end{equation}

	The symbol-wise AIR $\Rsym$ is achievable for a decoder that assumes \auxch. For the practical bit-interleaved coded modulation schemes \cite{Caire1998TransIT_BICM} that are also used in fiber-optics \cite{Alvarado2015JLT_BitWise}, a bit-wise demapper is followed by a binary decoder, as shown in Fig.~\ref{fig:model}. In this setup, the symbol-wise input \X is considered to consist of $m$ bit levels $\B=B_1,\dots,B_m$\footnote{A binary-reflected Gray code is used as labeling rule because it gives high BMD rates and offers the symmetry that is required for the shaping coded modulation scheme \cite[Sec.~IV-A]{Boecherer2015TransComm_ProbShaping}.} that can be stochastically dependent, and the decoder operates on bit-wise metrics. An AIR for this bit-metric decoding (BMD) scheme is the BMD rate \cite{Boecherer2016Arxiv_BMDRates}
	\begin{align}\label{eq:bmdrate} 
		\Rbmd & \triangleq \Biggl[ \sum_{i=1}^m \I(B_i;\Y) + \underbrace{ \Entr(\B) - \sum_{i=1}^m \Entr(B_i) }_{(\star)} \Biggr]^+ \\
		& = \Bigl[ \Entr(\B) - \sum_{i=1}^m \Entr(B_i|\Y) \Bigr]^+,
	\end{align}
	which is the AIR considered for the simulations in this work. In \eqref{eq:bmdrate}, the index $i$ indicates the bit level, ${\Entr}[\cdot]$ denotes entropy and $[\;\cdot\;]^+$ is $\max(\cdot\;,0)$. Note that \Rbmd is bounded above by the symbol-wise MI, $\I(\X;\Y) \geq \Rbmd$ \cite{Boecherer2016Arxiv_BMDRates}. The first term of \eqref{eq:bmdrate} is the sum of the MIs of $m$ parallel bit-wise channels. The term $(\star)$ in \eqref{eq:bmdrate} corrects for a rate overestimate due to dependent bit levels. For independent bit levels, i.e., $P_{\B}=\prod_{i=1}^m P_{B_i}$, the term $(\star)$ is zero and \Rbmd becomes the well-known generalized mutual information calculated with soft metrics that are matched to the channel. We calculate \Rbmd, which is an instance of \eqref{eq:mismatcheddecoderMI}, in Monte Carlo simulations of $N$ samples as
	\begin{align}\label{eq:bmdrate_MonteCarlo} 
		\Rbmd \approx & \frac{1}{N} \sum_{k=1}^{N} \left[ -\log_2 P_X(x_k) \right]\nonumber \\
		& -  \frac{1}{N} \sum_{k=1}^{N}\sum_{i=1}^m \left[\log_2\left( 1+e^{(-1)^{b_{k,i}} \Lambda_{k,i} }\right)\right],
	\end{align}
	where $b_{k,i}$ are the sent bits. The AIR \Rbmd is a function of the soft bit-wise demapper output $\Lambda_{k,i}$. These log-likelihood ratios (LLRs) are computed with the auxiliary channel as
	\begin{align}
		\Lambda_{k,i} & = \log \frac{\sum_{\x \in \setX_1^i} q_{\Y|\X}{(\y_k|\x)} P_X(x)}
			{\sum_{\x \in \setX_0^i}	 q_{\Y|\X}{(\y_k|\x)} P_X(x)}\\
		 & = \log \frac{q_{\Y|B_i}{(\y_k|1)}}{q_{\Y|B_i}{(\y_k|0)}} + \log \frac{P_{B_i}(1)}{P_{B_i}(0)}, \label{eq:defLLR}
	\end{align}
where $\setX_1^i$ and $\setX_0^i$ denote the set of constellation points whose $i$\textsuperscript{th} bit is 1 and 0, respectively. The first term of \eqref{eq:defLLR} is the LLR from the channel and the second term \tf{is} the a-priori information. For uniformly distributed input, the \tf{a-priori information is 0}. Using the 2D Gaussian auxiliary channel of \eqref{eq:auxchannel}, we have
	\begin{align} \label{eq:calcLLR}
		\Lambda_{k,i} & = \log \frac{\sum_{\x \in \setX_1^i}  e^{ -\frac{|\y_k-\x|^2}{2 \sigma^2} }P_X(x)}
			{\sum_{\x \in \setX_0^i}	e^{ -\frac{|\y_k-\x|^2}{2 \sigma^2}}P_X(x) }.
	\end{align}
These LLRs can be computed equivalently in 1D if a symmetric auxiliary channel is chosen, a product labeling is used \cite[Sec.~2.5.2]{Alvarado2015BICM_Book} and $\setX$ is generated from the product of 1D constellations.

\subsection{Probabilistic Shaping with Maxwell-Boltzmann PMFs}\label{ssec:shaping_with_MB}

	We search for the input distribution $P_{\X}^{*}$ that maximizes \Rbmd of \eqref{eq:bmdrate},
	\begin{equation}\label{eq:Shaping_optProblem}
		P_{\X}^{*} = \underset{P_{\X}:~\E[|X|^2]\leq 1}{\max}~ \Rbmd,
	\end{equation}	
	where the underlying channel is AWGN. Probabilistic shaping for the nonlinear fiber channel is discussed in Sec.~\ref{sec:mismatched_shaping_fiber}. As the AWGN channel is symmetric, the 1D PMFs are also symmetric around the origin, i.e.,
	\begin{equation}\label{eq:symmetric_PMF}
		P^{*}_{\text{1D}}(x) = P^{*}_{\text{1D}}(-x),
	\end{equation}
	which in 2D corresponds to a fixed probability per QAM ring. A common optimized input for \eqref{eq:Shaping_optProblem} is to use shaped input distributions from the family of Maxwell-Boltzmann (MB) distributions \cite[Sec.~IV]{Kschischang1993TransIT_Shaping}, \cite[Sec.~VIII-A]{Wachsmann1999TransIT_Shaping}. The method to find an optimized input for a particular SNR is discussed in detail in \cite[Sec.~III-C]{Boecherer2015TransComm_ProbShaping} and briefly reviewed in the following \tf{paragraph}.

	\begin{figure}[t]
		\centering
		\includegraphics{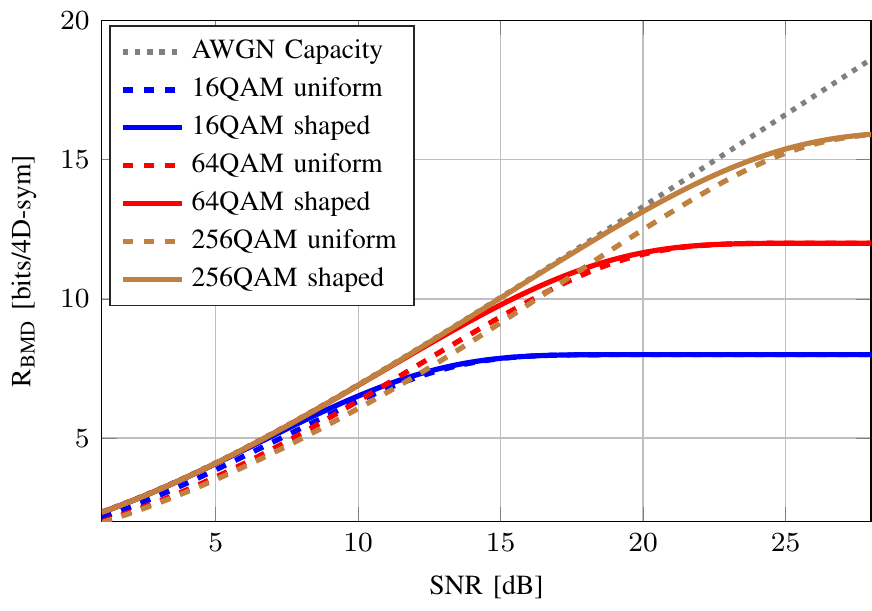}
		\caption{\Rbmd in bits/4D-sym for uniform QAM input (dashed lines) and QAM with the shaped, SNR-dependent MB PMF of \eqref{eq:MB_PMF} (solid lines). The AWGN capacity (dotted line) is shown as a reference.}
		\label{fig:BMD_AWGN_UniformShaped}
		\vspace{-7pt}
	\end{figure}

	\begin{figure*}[t]
		\centering
		\includegraphics{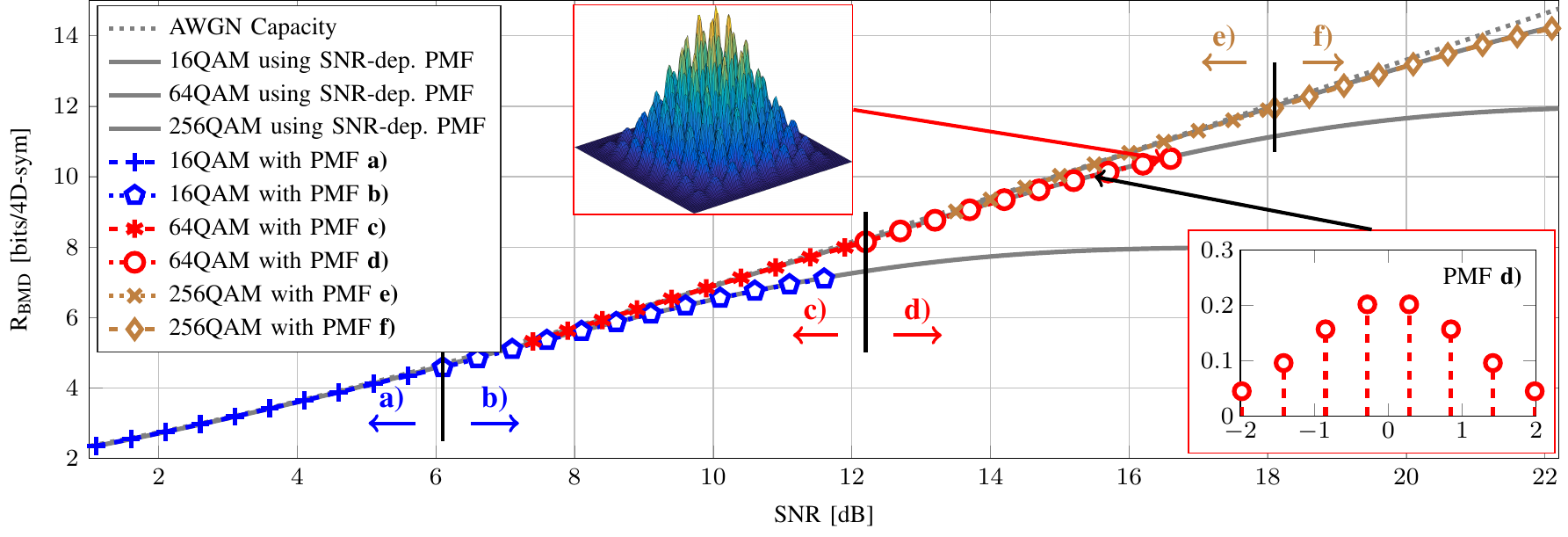}
		\caption{\Rbmd in bits/4D-sym for 16QAM, 64QAM, and 256QAM over the AWGN channel. The QAM formats shaped with SNR-dependent PMFs are shown as reference (gray solid lines), and shaping with fixed PMFs \textbf{a)} to \textbf{f)} is shown as colored lines with markers, with the respective SNR intervals indicated by vertical black lines. The inset on the bottom right shows the fixed PMF \textbf{d)} that is used for SNRs from 12.2~dB to 16.6~dB. \tf{The inset on the top shows the constellation diagram for PMF \textbf{d)} at 16.6~dB SNR}. Details on the fixed distributions are given in Table~\ref{table:input_dists_perfectshaping}.}
		\label{fig:BMD_MismatchedShaping_AWGN}
		\vspace{-7pt}
	\end{figure*}

	Let the positive scalar $\rho$ denote a constellation scaling of $\X$ with a fixed constellation $\setX$. Furthermore, let the PMF of the input be 
	\begin{align}\label{eq:MB_PMF}
		P^{*}_{\X}(\x_i)=\frac{1}{\sum_{j^{\prime}=1}^{M} e^{-\nu |\x_{j^{\prime}}|^2}}e^{-\nu |\x_i| ^2},
	\end{align}
	where $\nu$ is another scaling factor. For each choice of $\nu$, there exists a scaling $\rho$ that fulfills the average-power constraint $\E[|\rho X|^2]=1$. We optimize the scalings $\rho$ and $\nu$ such that \Rbmd is maximized while using a distribution from \eqref{eq:MB_PMF} and operating at the channel SNR that is defined as
	\begin{align}\label{eq:SNRconstraint_shaping}
	\text{SNR} \triangleq \frac{\E[|\rho X|^2]}{\sigma^2}=\frac{{E_\text{s}}}{N_0}=\frac{1}{N_0},
	\end{align}
	where the 1D signal power $E_\text{s}$ is normalized to 1 due to the average-power constraint and $N_0=\sigma^2$ is the noise variance per dimension. This optimization can be carried out with efficient algorithms, see \cite[Sec.~III-C]{Boecherer2015TransComm_ProbShaping}.

	In Fig.~\ref{fig:BMD_AWGN_UniformShaped}, \Rbmd in bits per four-dimensional symbol (bit/4D-sym) is shown versus the SNR of an AWGN channel. We choose to plot \Rbmd per 4D symbol to have values that are consistent with the dual-polarization AIRs of Sec~\ref{sec:mismatched_shaping_fiber}. Further, only \Rbmd is shown as it virtually achieves \Rsym \cite[Table~3]{Boecherer2015TransComm_ProbShaping}, \cite[Fig.~1]{Boecherer2016Arxiv_BMDRates}, and \Rbmd is the more practical AIR compared to \Rsym. The dotted curves represent uniformly distributed input and the solid lines show the \Rbmd for QAM with optimized MB input. The AWGN capacity $2\cdot\log_2(1+\text{SNR})$ is given for reference. Significant gains are found from probabilistic shaping over uniform input, with sensitivity improvements of up to 0.43~dB for 16QAM, 0.8~dB for 64QAM and more than 1~dB for 256QAM.

	\subsection{Shaping with Fixed PMFs}\label{ssec:bmd_awgn}
	
	In order to find the optimized MB input, the SNR of the channel over which we transmit, denoted \textit{channel} SNR, must be known or estimated \emph{a priori} at the transmitter. This transmitter-side estimate of the SNR is referred to as \textit{shaping} SNR. In a realistic communication system, it can be difficult to know the channel SNR at the transmitter because of varying channel conditions such as the number and properties of co-propagating signals, DSP convergence behavior, and aging of components. Hence, shaping without knowledge of the channel SNR could simplify the implementation of probabilistic shaping.
	We will see later that an offset from the shaping SNR to the channel SNR has a minor effect on \Rbmd for the AWGN channel if a suitable combination of QAM format and shaping SNR is used in the proper SNR regime.

	\begin{table}
	  \caption{Fixed distributions of $M$-QAM (with $E_s=1$) that lead to at most 0.1 \textrm{\normalfont dB} SNR loss compared to the full shaping gain}
    \setlength\tabcolsep{3pt}
    \vspace{-5pt}
    \centering
    \begin{tabular}{c||c|c||c|c||c|c}
    ~ & \textbf{a)} & \textbf{b)} & \textbf{c)} & \textbf{d)} & \textbf{e)} & \textbf{f)}\\
    \hline
      $M$-QAM & 16 & 16 & 64 & 64 & 256 & 256\\
    \hline
      \begin{tabular}{@{}c} Channel SNR \\ range [dB] \end{tabular} & \begin{tabular}{@{}c} 1.1\\ to\\ {6.1} \end{tabular} & \begin{tabular}{@{}c} 6.1\\ to\\ 11.6 \end{tabular} & \begin{tabular}{@{}c} 7.4\\ to\\ {12.2} \end{tabular} & \begin{tabular}{@{}c} 12.2\\ to\\ 16.6 \end{tabular} & \begin{tabular}{@{}c} 13.5\\ to\\ {18.1} \end{tabular} & \begin{tabular}{@{}c} 18.1\\ to\\ 22.1 \end{tabular}\\
    \hline
      Shaping SNR [dB] & 1.2 & 9.9 & 9.3 & 15.0 & 15.5 & 20.6\\
    \hline
      \begin{tabular}{@{}c} One-sided 1D PMF \\[1em] $\begin{bmatrix}P_{\text{1D}}(x_{\sqrt{M}/2+1})\\ \vdots \\ P_{\text{1D}}(x_{\sqrt{M}})\end{bmatrix}$ \end{tabular}
        &
          $\begin{matrix} 
            0.432\\
            0.068\\
          \end{matrix}$
        &
          $\begin{matrix} 
            0.332\\
            0.168\\
          \end{matrix}$
        &
          $\begin{matrix} 
            0.277\\
            0.16\\
            0.053\\
            0.01\\
          \end{matrix}$
        &
          $\begin{matrix} 
            0.200\\
            0.157\\
            0.096\\
            0.046\\
          \end{matrix}$
        &
          $\begin{matrix} 
            0.152\\
            0.131\\
            0.097\\
            0.062\\
            0.034\\
            0.016\\
            0.006\\
            0.002\\
          \end{matrix}$
        &
          $\begin{matrix} 
            0.109\\
            0.101\\
            0.088\\
            0.071\\
            0.054\\
            0.038\\
            0.025\\
            0.015\\   
          \end{matrix}$\\
    \hline
    \begin{tabular}{@{}c} One-sided 1D const. \\[1em] $\begin{bmatrix}x_{\sqrt{M}/2+1}\\ \vdots \\ x_{\sqrt{M}}\end{bmatrix}$ \end{tabular}
        &
          $\begin{matrix} 
            0.691\\
            2.072\\
          \end{matrix}$
        &
          $\begin{matrix} 
            0.521\\
            1.562\\
          \end{matrix}$
        &
          $\begin{matrix} 
            0.377\\
            1.131\\
            1.884\\
            2.638\\
          \end{matrix}$
        &
          $\begin{matrix} 
            0.282\\
            0.845\\
            1.409\\
            1.972\\
          \end{matrix}$
        &
          $\begin{matrix} 
            0.196\\
            0.588\\
            0.981\\
            1.373\\
            1.765\\
            2.157\\
            2.55\\
            2.942\\
          \end{matrix}$
        &
          $\begin{matrix} 
            0.147\\
            0.441\\
            0.734\\
            1.028\\
            1.322\\
            1.616\\
            1.909\\
            2.203\\
          \end{matrix}$\\
      \end{tabular}
    	\label{table:input_dists_perfectshaping}
	  \vspace{-10pt}
	\end{table}

	To realize large shaping gains, we choose to operate each QAM format within 0.1~dB of the AWGN capacity. The maximum SNRs under this constraint are found to be 6.1~dB for 16QAM, 12.2~dB for 64QAM, and 18.1~dB for 256QAM and depicted in Fig.~\ref{fig:BMD_MismatchedShaping_AWGN} as black vertical lines. We search numerically for the MB PMFs (obtained for a particular shaping SNR) that have at most 0.1~dB SNR loss compared to the full gain obtained when channel SNR and shaping SNR are identical. There are many distributions that fulfill this requirement, and we use the PMF that covers the largest SNR range while not exceeding the 0.1~dB penalty limit.
	The resulting PMFs of this numerical optimization are given as \textbf{a)}, \textbf{c)}, and \textbf{e)} in Table~\ref{table:input_dists_perfectshaping}\footnote{To reduce the size of Table~\ref{table:input_dists_perfectshaping}, we used the symmetry property of \eqref{eq:symmetric_PMF} and show the 1D PMF $P_{\text{1D}}$ for positive PAM constellations points $x_{\sqrt{M}/2+1},\dots,x_{\sqrt{M}}$ only.} and Fig.~\ref{fig:BMD_MismatchedShaping_AWGN}. We observe that a large SNR range is covered by a single PMF per QAM. However, these intervals are disconnected, and an additional PMF per QAM is necessary to cover the entire SNR range without gaps. 

	\tf{These additionally required shaped PMFs are given in Table~\ref{table:input_dists_perfectshaping} as PMFs \textbf{b)}, \textbf{d)}, and \textbf{f)}. Their operating range begins at the upper limit of the SNR range of \textbf{a)}, \textbf{c)}, and \textbf{e)}, while the upper limit is chosen to be at most 0.1~dB SNR below the full shaping gain obtained by an optimization for every SNR. This comes at the expense at operating away from capacity, and their gap to AWGN capacity is always larger than 0.1~dB.} The values of \Rbmd for all PMFs of Table~\ref{table:input_dists_perfectshaping} are shown in Fig.~\ref{fig:BMD_MismatchedShaping_AWGN}. We see that just two input distributions per modulation format are required to obtain a large shaping gain. \tf{This limited number of shaped input PMFs potentially allows an easier implementation of shaping when the rate gain from shaping can be utilized by a rate-adaptive FEC scheme.} In the remainder of the paper, we will investigate the impact of probabilistic shaping on fiber nonlinearities.

\section{SPM-XPM Model for the Nonlinear Fiber Channel}\label{sec:xpmmodel}

	\tf{The effective SNR, \SNR, of a signal after propagation over an optical fiber channel and receiver DSP is given by} \cite[Sec.~VI]{Poggiolini2014JLT_GNmodel},
	\begin{align}\label{eq:gnmodel_master}
		\SNR & = \frac{\Ptx}{\NoiseEff} = \frac{\Ptx}{\ASE + \NoiseNLI},
	\end{align}
	where \Ptx is the optical launch power, the noise term \ASE represents the amplified spontaneous emission (ASE) noise from the optical amplifiers and \NoiseNLI is the NLI \tf{variance} that includes both intra- and inter-channel distortions. In the classic GN model of \cite{Poggiolini2014JLT_GNmodel}, the nonlinearities are modeled as additive memoryless noise that follows a circularly symmetric (c.s.) Gaussian distribution. In particular, the choice of channel input \X in this model does not have an impact on \NoiseNLI in \cite[Sec.~VI]{Poggiolini2014JLT_GNmodel}, later shown in \cite{Dar2013OptExp_PropertiesNLIN} to be an inaccurate simplification. As a consequence, refined GN models have been presented in \cite{Dar2013OptExp_PropertiesNLIN,Carena2014OptExp_EGNmodel} that now include properties of the channel input in the modeling of \NoiseNLI, resulting in a more accurate representation of modulation-dependent nonlinear effects. These models allow us to study probabilistic shaping for the fiber-optic channel without computationally expensive SSFM simulations.

	In this work, we use the frequency-domain model of Dar \textit{et al.} \cite[Sec.~III]{Dar2013OptExp_PropertiesNLIN}\cite{Dar2015JLT_XPMModelInvited} with both intra-channel effects, i.e., self-phase modulation (SPM), and inter-channel effects, i.e., cross-phase modulation (XPM), which we refer to as the \textit{SPM-XPM model}. Four-wave mixing has been found numerically to give a negligible contribution to the total NLI for the considered multi-span fiber setup and is thus omitted in the analysis. In the following, we give an overview of the model and refer the reader to \cite[Sec.~III]{Dar2013OptExp_PropertiesNLIN} for details and derivations.

\subsection{SPM-XPM Model}\label{ssec:xpm_model}

	By rearranging the results in \cite[Sec.~III]{Dar2013OptExp_PropertiesNLIN} and \cite[App.]{Dar2014OptExp_NLINMatlab}, the NLI variance \NoiseNLI in \eqref{eq:gnmodel_master} can be expressed as
	\begin{align}\label{eq:xpmmodel_moments}
		\NoiseNLI = \Ptx^3 \left[ \right. \chinomom & + ( \standmom4 - 2) \cdot \chiforth \nonumber \\
		& + ( \standmom4 - 2)^2 \cdot \chiforth^{\prime} + \standmom6 \cdot \chisixth \left. \right],
	\end{align}
	where \standmom4 and \standmom6 are standardized moments of the input \X that are discussed in Sec.~\ref{ssec:stand_mom}, and \chinomom, \chiforth, $\chiforth^{\prime}$, and \chisixth are real coefficients that represent the contributions of the fiber nonlinearities.\footnote{\tf{Note that we have rearranged the results of} \cite[Sec.~3]{Dar2013OptExp_PropertiesNLIN} \tf{such that our coefficients} \chinomom, \chiforth, $\chiforth^{\prime}$, and \chisixth \tf{of this paper do not directly correspond to the coefficients} $\chi_1$ and $\chi_2$ of \cite[Eq.~25]{Dar2013OptExp_PropertiesNLIN} \tf{and the results of} \cite[App.]{Dar2014OptExp_NLINMatlab}. \tf{The rearranging allows to present the relation between coefficients and the moments} \standmom4 and \standmom6 \tf{more clearly}.} \tf{An implicit assumption of \eqref{eq:xpmmodel_moments} is that all WDM channels use the same modulation format and transmit at the same average launch power \Ptx, which is the case we consider throughout this paper.} Combining \eqref{eq:gnmodel_master} and \eqref{eq:xpmmodel_moments}, the total noise variance \NoiseEff is
		\begin{align}\label{eq:noiseff_total_modDependent}
		\NoiseEff = & \underbrace{\ASE + \Ptx^3 \cdot \chinomom}_\text{modulation-independent} \nonumber \\
		& + \underbrace{\Ptx^3 \left[  (\standmom4 - 2) \cdot \chiforth + ( \standmom4 - 2)^2 \cdot \chiforth^{\prime} + \standmom6 \cdot \chisixth \right]}_\text{modulation-dependent},
	\end{align}
  where we have split the overall noise into two terms. A modulation-independent noise contribution, given in the first line of \eqref{eq:noiseff_total_modDependent}, models ASE and partly NLI, and it is based solely on the system and fiber parameters, but not on the channel input. These two noise contributions are included in the classic GN model \cite[Sec.~VI]{Poggiolini2014JLT_GNmodel}. The expression in the second line of \eqref{eq:noiseff_total_modDependent} is a function of \momforth and \momsixth that are functions of the channel input, and thus models the modulation-dependency of \NoiseEff.

	\subsection{Standardized Moments}\label{ssec:stand_mom}

	We have seen that \NoiseEff in \eqref{eq:noiseff_total_modDependent} depends on the standardized moments \standmom4 and \standmom6. In general, the \kth standardized moment $\standmom{k}$ of the channel input \X is defined as
	\begin{equation}\label{eq:standmom_k_general}
	\standmom{k} = \frac{\E[|\X-\E[\X]|^k]}{(\E[|\X-\E[\X]|^2])^\frac{k}{2}} = \E[|\X|^k],
	\end{equation}
	where the final equality in \eqref{eq:standmom_k_general} holds because \X is symmetric around the origin (see \eqref{eq:symmetric_PMF}), which gives $\E[\X]=0$, and \X is normalized to unit energy, i.e., $\E[|\X|^2]=1$. 
	\begin{table}
	\caption{Overview of \standmom4 and \standmom6 of \eqref{eq:standmom_k_general} for different complex (2D) modulation formats and distributions}
	\label{table:overview_moments}
	\vspace{-5pt}
	\centering
	\begin{tabular}{r|c|c|c}
	Modulation & $P_{\X}$ & \standmom4 & \standmom6 \\
	\hline
	\hline
		 $M$-PSK &  uniform & 1 & 1 \\
		 16QAM &  uniform & 1.32 & 1.96 \\
		 64QAM &  uniform &  1.381 & 2.226 \\
		 256QAM &  uniform & 1.395 & 2.292 \\
 		 Continuous 2D & uniform & 1.4 & 2.316 \\
 		 \hline
		 16QAM &  \textbf{b)} of Table~\ref{table:input_dists_perfectshaping}&  1.525 & 2.76 \\
		 64QAM &  \textbf{d)} of Table~\ref{table:input_dists_perfectshaping} &  1.664& 3.518 \\
		 256QAM &  \textbf{f)} of Table~\ref{table:input_dists_perfectshaping} &  1.713 & 3.808 \\
		 \hline
 		 Continuous 2D & Gaussian &  2 & 6 \\
		\end{tabular}
	\vspace{-10pt}
	\end{table}

	Table~\ref{table:overview_moments} shows the moments \standmom4 and \standmom6 for different modulation formats and PMFs. Constant-modulus modulation such as phase-shift keying (PSK) minimizes both moments. For uniform QAM, \standmom4 and \standmom6 increase with modulation order. The limit for complex uniform input is given by uniform QAM with infinitely many signal points, which corresponds to a continuous uniform input in 2D. When the input is shaped with an MB PMF that fulfills \eqref{eq:Shaping_optProblem}, e.g., with the fixed distributions in Table~\ref{table:input_dists_perfectshaping}, \standmom4 and \standmom6 are larger than for a uniform input distribution. A complex continuous Gaussian density gives the respective maxima for these moments.

	\subsection{NLI Increase due to Shaping}

	The modulation-dependent coefficients \chiforth, $\chiforth^{\prime}$, and \chisixth in \eqref{eq:noiseff_total_modDependent} together with the results in Table~\ref{table:overview_moments} give us an insight into how the choice of a particular modulation affects \NoiseNLI. Considering the first modulation-dependent term in \eqref{eq:noiseff_total_modDependent}, a small \standmom4 corresponds to a decrease in $(\standmom4-2)$ and thus, to less NLI. PSK formats, for example, minimize \standmom4 and \standmom6 and thus induce a minimum amount of NLI, which is why these formats can have superior performance to QAM in cases \tf{in which the contribution of \chiforth is significant, e.g. in single-span links or} systems with inline dispersion management \cite[Fig.~4]{KoikeAkino2016OFC_ModulationDispManagement}. On the other hand, distributions that are well-suited for the AWGN channel, such as the shaped PMFs in Table~\ref{table:overview_moments}, have increased moments \standmom4 and \standmom6, which results in stronger NLI than uniform input. 
	The interpretation of probabilistic shaping for the nonlinear fiber channel is then that a shaping gain can be obtained from the channel portion described by the linear noise contribution \ASE and the moment-independent term $\chinomom$. Simultaneously, an increase in NLI is introduced by shaping as the modulation-dependent term of \eqref{eq:noiseff_total_modDependent} becomes larger, and the optimal trade-off is not obvious. \tf{This behavior has previously been investigated in} \cite{Dar2014ISIT_nonlinearShaping} \tf{for multi-dimensional constellations and short fiber links, for which shaping gains of more than 1.53~dB were found. We will numerically study this trade-off between shaping gain and shaping penalty for QAM in detail in Sec.~\ref{ssec:shaping_sensitivity}.}

\section{Probabilistic Shaping of 64QAM for a Multi-Span Fiber System}\label{sec:mismatched_shaping_fiber}
	In the following, we numerically evaluate the AIRs for a multi-span fiber link with uniform and shaped 64QAM input. The analysis focuses on the effect of shaping on the fiber nonlinearities and therefore, on the effective SNR. Transceiver impairments and further effects that require \tf{or result from} advanced DSP are not included in this work.

\subsection{Fiber Simulations}\label{ssec:sim_setup}
	
	\begin{table}
	\caption{System and Simulation Parameters}
	\label{table:sim_params}
	\vspace{-5pt}
	\centering
	\begin{tabular}{c|c}
		Parameter & Value \\
		\hline
		\hline
		Modulation & 64QAM \\
		Input PMF & uniform and shaped \\
		Polarization & dual-polarization \\
		Symbol rate & 28 GBaud \\
		Pulse shape & root-raised cosine (RRC) \\
		RRC roll-off & 0.01 \\
		WDM channels & 9 \\
		WDM spacing & 30 GHz \\
		Nonlinear coefficient $\gamma$ & 1.3 1/W/km \\
		Dispersion & 17 ps/nm/km \\
		Attenuation $\alpha$ & 0.2 dB/km \\
		Length per span & 100 km \\
		Amplification & EDFA \\
		EDFA noise figure & 4 dB \\
		Dispersion compensation & digital \\
		Demapper statistics & c.s. Gaussian \\
		\hline 
		SSFM step size & 0.1 km \\ 
		Oversampling factor & 32 \\ 
		QAM symbols \tf{per WDM ch.} (SSFM sim.) & 500,000 \\
		Samples (SPM-XPM model) & 1,000,000 \\
		\end{tabular}
	\vspace{-10pt}
	\end{table}

	A multi-span optical fiber system is simulated, with the main parameters given in Table~\ref{table:sim_params}. 64QAM symbols are generated with the constant-composition distribution matcher \cite{Schulte2016TransIT_DistributionMatcher} and shaped according to the specified input distribution. Pulse shaping is done digitally and the resulting signal is transfered ideally into the optical domain. The center WDM channel is the channel of interest and all WDM channels have the same PMF, but with \tf{statistically independent symbol sequences}. The propagation of the signal over each span fiber is simulated using the SSFM. After each span, the signal is amplified by an EDFA and ASE noise is added. At the receiver, the center WDM channel is filtered and ideally transferred into the digital domain. Chromatic dispersion is digitally compensated for, a matched filter is applied, and the signal is downsampled. The channel SNR is computed as the average over both polarizations, and \Rbmd per polarization is calculated as stated in \eqref{eq:bmdrate_MonteCarlo} and we sum over both polarizations. For the BMD rate estimation, 2D c.s. Gaussian statistics as given in \eqref{eq:auxchannel} \cite{Fehenberger2015OptExp_AIR} with static mean values\footnote{For static mean values, the centroids of the Gaussian distributions are identical to the sent constellation points $\x$, as stated in \eqref{eq:auxchannel}. In contrast, using adaptive mean values \cite{Eriksson2016JLT_4DMI} means that the centroids are calculated from the received symbols $\y_k$.} are used.

\subsection{Numerical Evaluation of the SPM-XPM Model}\label{ssec:xpmmodel_sims}

	\begin{table}
	\caption{ASE noise and NLI terms in \eqref{eq:noiseff_total_modDependent}}
	\label{table:xpmmodel_chi_values_noPower}
	\vspace{-5pt}
	\centering
	\begin{tabular}{c|c}
		Noise term & Value \\
		\hline
		\hline
		\ASE & 17.85\tento{-6} W \\
		$\chinomom$ & 3.09\tento4 W\textsuperscript{-2} \\
		$\chiforth$ & 1.05\tento4 W\textsuperscript{-2} \\
		$\chiforth^{\prime}$ & -1.22 \tento2 W\textsuperscript{-2} \\
		$\chisixth$ & 1.29 \tento2 W\textsuperscript{-2} \\
		\end{tabular}
	\vspace{-10pt}
	\end{table}
 
  The nonlinear terms \chinomom, \chiforth, $\chiforth^{\prime}$, and \chisixth in \eqref{eq:noiseff_total_modDependent} are calculated via Monte Carlo simulations for which a ready-to-use web interface \cite{NLINwizard} and MATLAB code \cite[App.]{Dar2014OptExp_NLINMatlab} is available. Note that \eqref{eq:noiseff_total_modDependent} includes inter-channel and intra-channel effects as well as additional intra-channel terms that occur at very dense WDM spacings, as discussed in \cite{Carena2014OptExp_EGNmodel,NLINwizard}. We also note that virtually identical results are obtained for the sinc pulse shape described in \cite[App.]{Dar2014OptExp_NLINMatlab} and the narrow RRC filtering in this work.

	For the parameters given in Table~\ref{table:sim_params} and a transmission distance of 2,000~km, the amplifier noise power \ASE and the nonlinear coefficients of \eqref{eq:noiseff_total_modDependent} are given in Table~\ref{table:xpmmodel_chi_values_noPower}. We observe that \chinomom and \chiforth are the dominant NLI contributions. The values of all NLI terms are used to compute the effective \SNR of \eqref{eq:gnmodel_master}. The BMD rate in \eqref{eq:bmdrate} is computed by numerical integration. Note that $\chiforth^{\prime}$ is negative for the considered system parameters. In this case, we have a negative term $(\standmom4 - 2)^2 \cdot \chiforth^{\prime}$ and thus, a larger \NoiseNLI for an increased \standmom4, which is the same behavior that is observed for the term $(\standmom4-2) \cdot \chiforth $.

\subsection{Reach Increase from Shaping}\label{ssec:sim_reach}

  \begin{figure}[t]
		\includegraphics{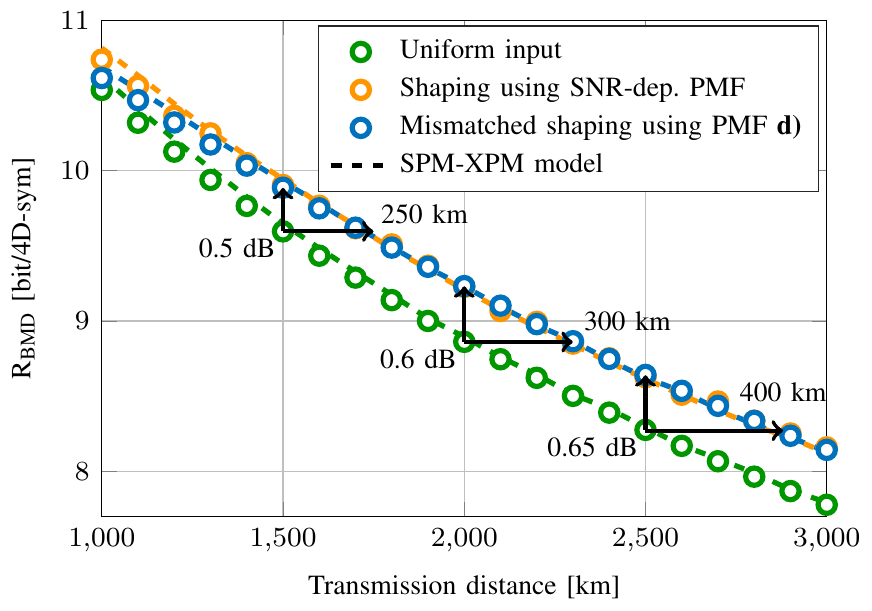}
  	\caption{\Rbmd in bit/4D-sym versus transmission distance in km for the SPM-XPM model of Sec.~\ref{sec:xpmmodel} (dashed lines) with 64QAM input and SSFM simulations (markers). The colors represent the different PMFs: uniform input (green), SNR-dependent shaped PMFs (orange), and the SNR-independent PMF \textbf{d)} of Table~\ref{table:input_dists_perfectshaping} (blue).}
		\label{fig:SSFM_distanceSweep_UniformMatchedMismatched}
  \end{figure}

	We compare \Rbmd for 64QAM with uniform input, with an MB input PMF that is dependent on the transmission distance (and thus on the channel SNR, see Sec.~\ref{ssec:shaping_with_MB}), and with the fixed PMF \textbf{d)} of Table~\ref{table:input_dists_perfectshaping}. Figure \ref{fig:SSFM_distanceSweep_UniformMatchedMismatched} shows \Rbmd in bit/4D-sym for transmission distances from 1,000~km to 3,000~km in steps of 100~km. Results for SSFM simulations (markers) and for the SPM-XPM model (dashed lines) are shown, and we observe a good agreement between them. For each transmission distance, the launch power is varied with a granularity of 0.5~dB and the optimal power is used, which is $-$1.5~dBm or $-$1~dBm per WDM channel for all distances and input PMFs.

	The channel SNR for uniform 64QAM is between 17.2~dB SNR for 1,000~km and 12.35~dB SNR for 3,000~km, and the SNR for each distance is used as the shaping SNR for the SNR-dependent input PMF. Using this shaped input gives an AIR gain over uniform input for a fixed transmission distance or, equivalently, an increase in transmission distance for a fixed AIR. For example, shaping gives a 300~km reach improvement, from 2,000~km to 2,300~km, at an AIR of 8.86~bit/4D-sym. Similar gains are observed for all link lengths and are in agreement with previous shaping simulations of a WDM system \cite{Fehenberger2015OFC_ProbShaping}\cite[Sec.~3.5]{Fehenberger2015OptExp_AIR}. The AIR gains from shaping translate to sensitivity improvements of up to 0.65~dB, which is slightly below the maximum shaping gain of 0.8~dB seen for 64QAM in back-to-back experiments \cite[Fig.~2]{Fehenberger2016PTL_MismatchedShaping} where no NLI is present, and larger gains are possible with higher-order modulation. For distances between 1,400~km and 3,000~km, the PMF \textbf{d)} gives identical gains to those of the shaped input that is matched to the SNR at each transmission distance. For smaller distances, a gap between mismatched and SNR-dependent shaping exists because the system is operated in the high-SNR regime beyond the channel SNR range of PMF \textbf{d)}, in which case switching to 256QAM is advisable.

\subsection{AIR Gain of Shaped 64QAM at 2,000~km Distance}\label{ssec:sim_2000km}
	\begin{figure}[t]
		\includegraphics{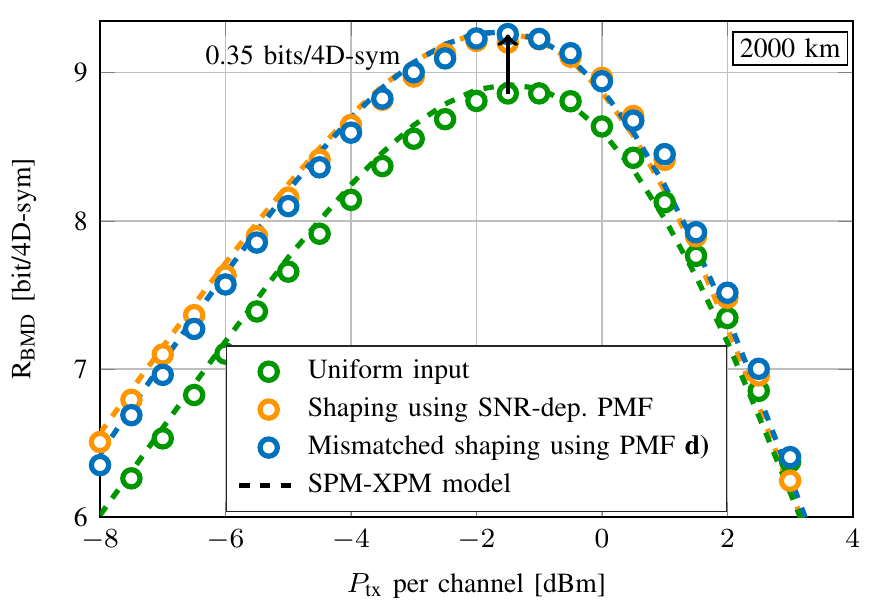}
		\caption{\Rbmd in bit/4D-sym versus \Ptx per channel in dBm for 64QAM. The SPM-XPM model results (dashed lines) for uniform input, SNR-dependent shaping and shaping with a fixed PMF are shown. For the latter, the distribution \textbf{d)} of Table \ref{table:input_dists_perfectshaping} is used for all launch powers. SSFM simulation results (markers) validate the SPM-XPM model.}
		\label{fig:SSFM_2000km_UniformMatchedMismatched_EGN_AIRvsPower}
	\end{figure}

	The effects of shaping in the presence of fiber nonlinearities are investigated for a transmission distance of 2,000~km (all other parameters are as given Sec.~\ref{ssec:sim_setup}). 
	In Fig.~\ref{fig:SSFM_2000km_UniformMatchedMismatched_EGN_AIRvsPower}, \Rbmd in bit/4D-sym is shown versus \Ptx per channel in dBm. A good match between simulation results (markers) and the SPM-XPM model (dashed lines) is again observed. At the optimum launch power, a shaped input distribution gives an AIR improvement of 0.35~bit/4D-sym over uniform input. For all relevant launch powers, SNR-dependent shaping and shaping with the fixed PMF \textbf{d)} give identical gains. This shows again that it is sufficient to use just one input distribution to realize the shaping gain for various transmit powers. We see from the SPM-XPM model that in the highly nonlinear regime, the shaping gain is significantly reduced and disappears for very high launch powers, which is due to the adverse ramifications of shaping. The sensitivity of these effects on \SNR and the AIRs is investigated next.

\subsection{Sensitivity Analysis of Probabilistic Shaping}\label{ssec:shaping_sensitivity}

	In the following, the sensitivity of NLI on probabilistic shaping is studied. The SNR mismatch between shaping SNR and channel SNR is chosen as a figure of merit for this analysis as it describes how strongly a QAM input is shaped with one single number that parameterizes an MB PMF. The SNR mismatch is denoted by $\Delta$ and calculated for each simulation run as the shaping SNR that is used at the transmitter minus the channel SNR that is estimated after the DSP. The chosen definition of $\Delta$ means that a larger $\Delta$, i.e., a large shaping SNR, corresponds to a distribution that is closer to uniform, while a smaller $\Delta$ represents a PMF that is strongly shaped. The mismatch $\Delta$ was varied by diverting from the channel SNR of uniform 64QAM at 2,000~km, which is 14.17~dB, in steps of 0.1~dB. These values are used as shaping SNRs. In total, 100 full SSFM simulation runs with a transmission length of 2,000~km were performed to gather sufficient statistics for $\Delta$ in the range of $-$4~dB to 6~dB. 

  \begin{figure}[t]
		\includegraphics{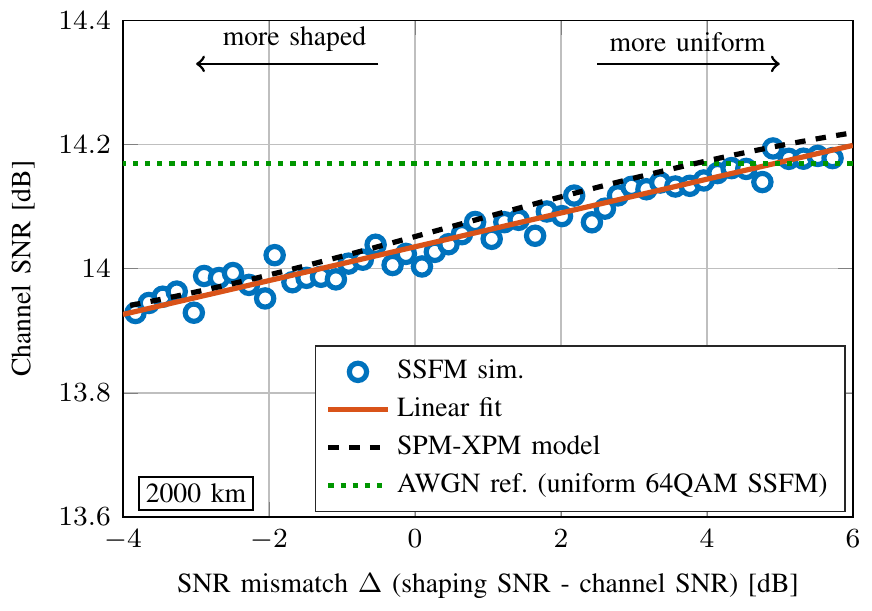}
  	\caption{Channel SNR measured after DSP versus the SNR mismatch $\Delta$, both in dB. The SPM-XPM model (dashed line) shows the dependence of the channel SNR on the input. This behavior would not be present in a linear AWGN channel (dotted line), which also represents the channel SNR for SSFM simulations with uniform 64QAM. Further SSFM simulations of shaped 64QAM over 2,000~km at the optimum launch power are also shown (circles), where each marker represents a simulation run. The solid orange line is a linear fit of the simulation results.}
		\label{fig:ChannelSNR_SNRMismatch_SSFM}
  \end{figure}

	\subsubsection{Shaping Decreases SNR}
	Figure~\ref{fig:ChannelSNR_SNRMismatch_SSFM} shows the dependence of the channel SNR on $\Delta$, with blue markers representing simulation results for shaped 64QAM over 2,000~km and the solid curve being a linear fit to the simulations. In the considered range of $\Delta$, we observe a good match of the simulation data to a linear fit. The results of the SPM-XPM model, shown as the dashed curve, are within 0.05~dB of the fit and are an accurate approximation of the simulation results. Hence, the SPM-XPM model correctly predicts the growth of \NoiseNLI, and thus the decrease of the effective SNR, with increasing moments \standmom4 and \standmom6 that result from a decrease in $\Delta$. \tf{We further observe that for large $\Delta$'s, i.e., virtually uniform input, the simulations results approach the SNR for uniform input (dotted line), and the fluctuations are due to the limited accuracy of Monte-Carlo simulations. The SPM-XPM model slightly overestimates the AIR by approximately 0.05 dB.}
	The AWGN reference (dotted line), which also corresponds to constant channel SNR of uniform 64QAM input in SSFM simulations, confirms that for a linear channel without NLI, the channel SNR does not depend on the input distribution. However, it is important to realize that the increase in SNR that is observed for increasing $\Delta$ does not imply a gain in AIR, as we will show next.

	\subsubsection{Shaping Increases AIR}
\begin{figure}[t]
		\includegraphics{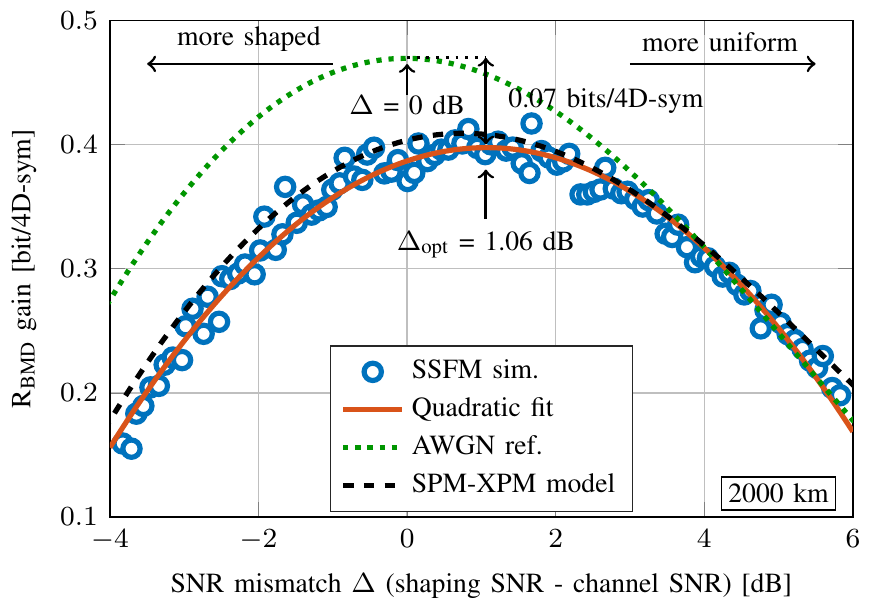}
  	\caption{\Rbmd gain in bit/4D-sym of shaped 64QAM over uniform 64QAM  versus SNR mismatch $\Delta$ in dB. The solid line is a quadratic fit to the simulations and has its maximum at $\Delta_\text{opt}$~=~1.06~dB. The SPM-XPM model (dashed line) correctly predicts the trend of the simulations.
  	The dotted line shows as a reference the BMD rate gain for an AWGN channel with 14.17~dB SNR where the maximum shaping gain is obtained in the case of zero SNR mismatch, i.e., $\Delta$~=~0~dB.}
		\label{fig:AIR_SNRMismatch_SSFM}
  \end{figure}

	In Fig.~\ref{fig:AIR_SNRMismatch_SSFM}, the sensitivity of the shaping gain is investigated by plotting the \Rbmd gain over uniform 64QAM (with an \Rbmd of 8.86~bit/4D-sym) as a function of the SNR mismatch $\Delta$. Blue markers indicate SSFM simulations and a quadratic fit to the simulation results is given as the solid line. The fitted parabola is relatively flat around its peak of 0.4~bit/4D-sym, and significant shaping gains of more than 0.3~bit/4D-sym are obtained for $\Delta$ from approximately -2~dB to 4~dB. Hence, only small penalties are to be expected in shaping gain in comparison to matched shaping if the SNR mismatch is within a range of several dB. This is again supported by the SPM-XPM model (dashed curve) that accurately predicts this behavior.

	Another interesting aspect of Fig.~\ref{fig:AIR_SNRMismatch_SSFM} is that the maximum shaping gain, according to the quadratic fit of the simulation data, is found at $\Delta_\text{opt}$~=~1.06~dB. In the absence of fiber nonlinearities, i.e., for an AWGN channel, zero mismatch, i.e., $\Delta$~=~0~dB, is expected to be optimal, as confirmed by the AWGN reference (dotted curve). As explained by the SPM-XPM model in Sec.~\ref{ssec:xpm_model} and also in the context of Fig.~\ref{fig:ChannelSNR_SNRMismatch_SSFM}, a shaped input causes stronger NLI than does a uniform one. Thus, $\Delta_\text{opt}$ is expected to be larger than 0~dB since a positive $\Delta$ indicates a more uniform-like input that introduces less NLI. However, the magnitude of this effect is very small and significant shaping gains are observed around the optimum SNR mismatch. The NLI increase due to shaping is also the reason for the difference in \Rbmd gain between the optical fiber simulations and the AWGN channel. This gap of approx. 0.07~bit/4D-sym at the optimum $\Delta$ disappears for large positive values of $\Delta$ because, in this range, a uniform input is approached and increased NLI due to shaping no longer occurs. The effect that causes the gap between AWGN and fiber simulations can be considered as a shaping penalty due to increased NLI, and we observe its magnitude to be relatively small.

\subsection{Optimized Shaping for the Nonlinear Fiber Channel}\label{ssec:shaping_nonlinear}
	So far, we have restricted our analysis of probabilistic shaping to distributions of the MB family, see \eqref{eq:MB_PMF}, and have considered only 1D PMFs that were then extended to 2D. We have shown that these inputs are an excellent choice for the AWGN channel, and that large shaping gains are also obtained for the optical channel. However, it is not clear whether we can find a better shaped input PMF for the nonlinear fiber channel because there might be input distributions that have a better trade-off between the shaping gain and the shaping penalty due an NLI increase.

	In the following, we use the SPM-XPM model of Sec.~\ref{ssec:xpm_model} as the channel for the optimization problem \eqref{eq:Shaping_optProblem} and numerically search for the shaped inputs that give the largest \Rbmd. \tf{This optimization is conducted in MATLAB using the interior point algorithm} \cite{Byrd2000MathProgramming_InteriorPoint}.	We consider inputs in 1D (which are then extended to 2D), and also optimize PMFs directly in 2D. This 2D approach gives us more degrees of freedom in the optimization problem and allows us to consider any probabilistically shaped input in 2D, including multi-ring constellations \cite[Sec.~IV-B]{Essiambre2010JLT_CapacityLimits}. However, the benefit of 1D PMFs is that the receiver can operate in 1D, without loss of information for a symmetric channel, which leads to reduced demapper complexity compared to the 2D case.

\begin{figure}[t]
    \includegraphics{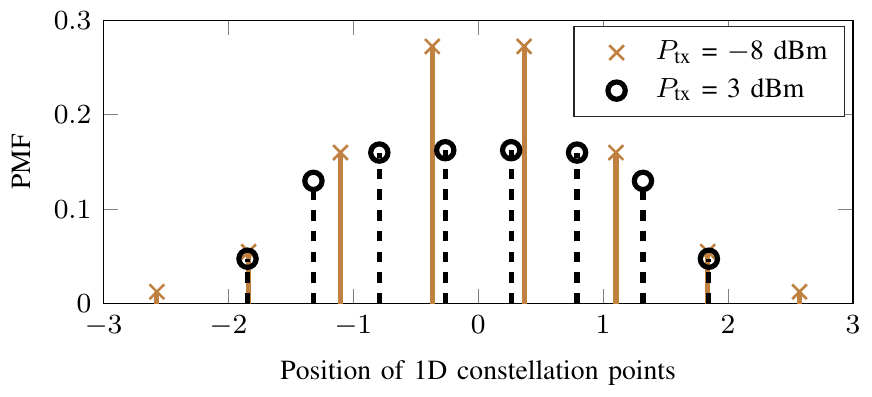}
  	\caption{Optimized 1D PAM PMFs for \Ptx~=~$-$8~dBm (solid, crosses) and \Ptx~=~3~dBm (dashed, circles). Although these two power levels have approximately the same \SNR, the PMF for 3~dBm is not shaped as much in order to avoid increased NLI.}
		\label{fig:opt_inputPMF}
  \end{figure}

	The system described in Sec.~\ref{ssec:sim_setup} with a distance of 2,000~km is considered for the optimization. In Fig.~\ref{fig:opt_inputPMF}, two 1D PMFs are shown that are the respective results of the optimization problem in the linear regime at $\Ptx =-$8~dBm and in the nonlinear regime at $\Ptx =$3~dBm. Details of the PMFs (and for completeness the PMFs at the optimum \Ptx) are given in Table~\ref{table:xpmmodel_opt_PMF}. Despite the two different launch powers, their effective SNRs \SNR are virtually identical, and using MB PMFs that are based solely on the channel SNR would result in the same PMF for both launch powers. This restriction is lifted in the optimization problem under consideration, and we observe from Fig.~\ref{fig:opt_inputPMF} that the PMF for 3~dBm is not shaped as much as the one for $-$8~dBm. This illustrates that strong shaping is avoided at high power levels when a PMF optimization with the SPM-XPM model is performed.

	\begin{table}[t]
	\caption{Input PMFs optimized with the SPM-XPM model}
	\label{table:xpmmodel_opt_PMF}
	\vspace{-5pt}
	\centering
	\begin{tabular}{c|c|c|c|c}
		\Ptx per ch. & One-sided 1D PMF & \standmom4 & \standmom6 & \SNR\\
		\hline
		\hline
		$-$8~dBm & [0.2725, 0.16, 0.055, 0.0125] &  1.918 & 5.145 & 9.46~dB\\
		3~dBm & [0.1625, 0.16, 0.13, 0.0475] & 1.511 & 2.822 & 9.49~dB\\
		\hline	
		$-$1.5~dBm & [0.195, 0.1625, 0.1, 0.0425] & 1.642 & 3.432 & 14.09~dB\\
		$-$1.5~dBm & 2D PMF: see Fig.~\ref{fig:EGN_Model_Optim_2000km} (inset) & 1.599 & 3.197 & 14.12~dB\\
		\end{tabular}
	\vspace{-10pt}
	\end{table}

\begin{figure}[t]
    \includegraphics{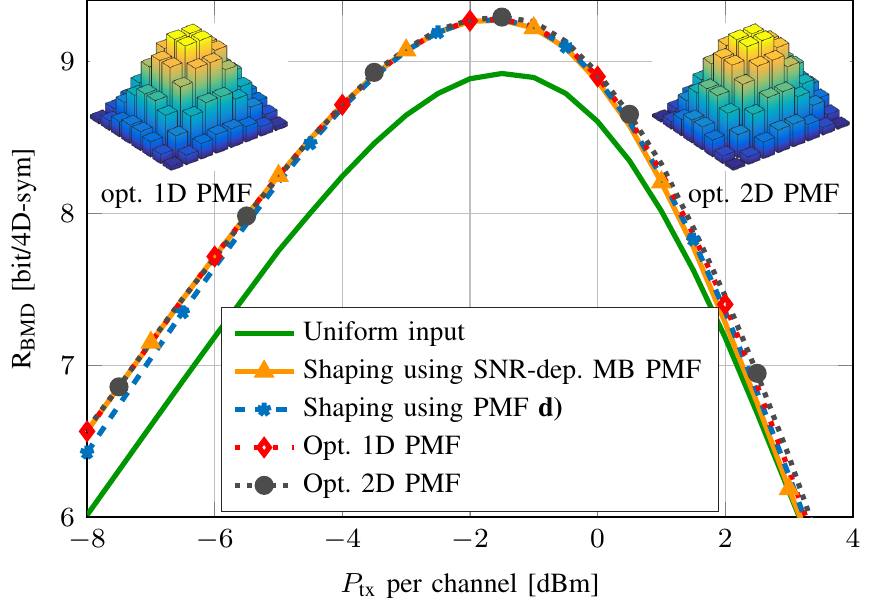}
  	\caption{\Rbmd in bit/4D-sym for 64QAM versus \Ptx per channel in dBm for the SPM-XPM model. The AIRs for the 1D-optimized input (red dotted), 2D-optimized PMF (gray dotted) and for all other shaped inputs lie on top of each other over a wide range of launch powers. Inset: The optimized 1D PMF in its 2D representation and the 2D PMF, each for \Ptx of $-$1.5~dBm.}
		\label{fig:EGN_Model_Optim_2000km}
  \end{figure}

	In Fig.~\ref{fig:EGN_Model_Optim_2000km}, \Rbmd is shown versus \Ptx per channel for different input distributions of 64QAM. All results are obtained from the SPM-XPM model. The dotted curves show \Rbmd for a 1D PMF (red) and a 2D PMF (gray), both optimized with the SPM-XPM model, and the AIRs for uniform, MB-shaped input, PMF \textbf{d)} are also included. The two optimized PMFs give identical values of \Rbmd, and their shapes are very similar, as the insets in Fig.~\ref{fig:EGN_Model_Optim_2000km} show. We conclude that, for the considered system, there is virtually no benefit in using the optimized 2D input. Additionally, the MB shaped input gives identical gains to the 1D-optimized input at low power and around the optimal one. It is only in the high-power regime that slightly increased AIRs are obtained with the optimized input. This indicates that, also for a multi-span fiber channel, the shaping gain is very insensitive to variations in the input distribution, and an optimized input gives shaping gains that are no larger than those with an MB PMF. In fact, it is sufficient for the considered system simply to use the fixed input distribution \textbf{d)} from Table~\ref{table:input_dists_perfectshaping} to effectively obtain the maximum shaping gain.

\section{Conclusions}
In this work, we have studied probabilistic shaping for long-haul optical fiber systems via both numerical simulations and a GN model. We based our analysis on AWGN results that show that just two input PMFs from the family of Maxwell-Boltzmann distributions are sufficient per QAM format to realize large shaping gains over a wide range of SNRs. We have found that these fixed shaped distributions also represent an excellent choice for applying shaping to a multi-span fiber system. Using just one input distribution for 64QAM, large shaping gains are reported from transmission distances between 1,400~km to 3,000~km. For a fixed distance of 2,000~km, we have studied the impact of probabilistic shaping with Maxwell-Boltzmann distributions and other PMFs. The adverse effects of shaping in the presence of modulation-dependent nonlinear effects of a WDM system have been shown to be present. An NLI penalty from shaping is found to be relatively minor around the optimal launch power in a multi-span system. This means that, for the considered system, just one input PMF for 64QAM effectively gives the maximum shaping gain and an optimization for the fiber channel is not necessary. This could greatly simplify the implementation and design of probabilistic shaping in practical optical fiber systems. We expect similar results for other QAM formats such as 16QAM or 256QAM when they are used in fiber systems that are comparable to the ones in this work. We have also found that the GN model is in excellent agreement with the SSFM results, confirming its accuracy for shaped QAM input. 

For nonlinear fiber links \tf{in which the contribution of \chiforth is significant}, e.g., those with in-line dispersion management or single-span links with high power, further optimizations of the shaping scheme can be both beneficial for a large shaping gain and incur low NLI. Additionally, instead of shaping on a per-symbol basis, constellation shaping over several time slots to exploit the temporal correlations by XPM is an interesting future step to increase SE. Also, optimizing distributions in four dimensions could be beneficial for highly nonlinear polarization-multiplexed fiber links.

\section{Acknowledgments}
The authors would like to thank Prof. Frank Kschischang (University of Toronto) for encouraging us to use the SPM-XPM model to study probabilistic shaping for the nonlinear fiber channel. \tf{The authors would also like to thank the anonymous reviewers for their valuable comments that helped to improve the paper.}

\end{document}